\documentclass[summary]{ursi}
\usepackage{amsmath}
\usepackage{amssymb}
\usepackage{algorithm}
\usepackage{algorithmic}

\title{Pinching-Antenna System (PASS)-enabled Multicast Communications}

\author{Xidong Mu\affref{ref1}, Guangyu Zhu\affref{ref2},
  and Yuanwei Liu\affref{ref3}}

\affiliation{%
  \aff{ref1}{Centre for Wireless Innovation, Queen's University Belfast, Belfast, BT3 9DT, U.K.}
  \aff{ref2}{School of Artificial Intelligence, Beijing University of Posts and Telecommunications, Beijing, 100876, China.}
  \aff{ref3}{Department of Electrical and Electronic Engineering, The University of Hong Kong, Hong Kong.}
}

\begin{document}

\maketitle

\begin{abstract}
Pinching-antenna system (PASS) is a novel flexible-antenna technology, which employs long-spread waveguides to convey signals with negligible path loss and pinching antennas (PAs) with adjustable positions to radiate signals from the waveguide into the free space. Therefore, short-distance and strong line-of-sight transmission can be established. In this paper, a novel PASS-enabled multicast communication framework is proposed, where multiple PAs on a single waveguide radiate the broadcast signals to multiple users. The multicast performance maximization problem is formulated to optimize the positions of all PAs. To address this non-convex problem, a particle swarm optimization-based algorithm is developed. Numerical results show that PASS can significantly outperform the conventional multiple-antenna transmission.
\end{abstract}

\section{Introduction}
Throughout the history of wireless communications, multiple-antenna technologies have been key enablers of performance enhancements. For example, massive multiple-input multiple-output (MIMO) has significantly improved communication efficiency through spatial multiplexing and beamforming gains~\cite{6736761}. More recently, researchers have turned their attention to flexible-antenna technologies, such as reconfigurable intelligent surfaces (RISs)~\cite{9140329}, fluid-antenna systems~\cite{9264694}, and movable antennas~\cite{10286328}. These emerging technologies represent a new paradigm for 6G and beyond, enabling the beneficial reconfiguration of wireless channels between transceivers and further enhancing communication performance. 

Among these innovations, the pinching-antenna system (PASS) has been proposed as a revolutionary technology in the family of flexible-antenna technologies~\cite{DOCOMO}. Generally speaking, PASS utilizes dielectric waveguides as the transmission medium, where signals experience significantly low propagation loss. Small dielectric elements, known as pinching antennas (PAs), are placed on the waveguide with adjustable activation positions to radiate signals into free space. By strategically distributing waveguides and carefully selecting the activation positions of PAs, PASS can establish short-distance line-of-sight (LoS) links, thereby providing high-quality wireless services to user terminals~\cite{PASS}. This approach mitigates the substantial power loss associated with long-distance propagation and effectively addresses signal blockage issues, particularly in mmWave and THz communications. Note that the simple structure of PASS makes it highly adaptable and easy to deploy in real-world scenarios, including indoor ceilings, building facades, and roadsides, among other environments. Therefore, PASS becomes a promising solution for next-generation wireless networks. For example, the authors of \cite{Ding} first initialized the theoretical performance analysis of employing PASS in communication systems, where both single-waveguide and multiple-waveguide cases were considered to derive the ergodic rate achieved by PASS. Moreover, the authors of \cite{Kaidi} studied a PASS-enabled non-orthogonal multiple access communication system, where the discrete PA activation positions were optimized under the consideration of in-waveguide propagation loss. Existing works have demonstrated the superiority of PASS over conventional MIMO.

Recall the fact that a single waveguide can only be fed with the same signal \cite{PASS,Ding,Kaidi}. This inherent characteristic makes PASS particularly well-suited for supporting multicast communications, where a common message must be delivered to multiple users (e.g., video streaming and public entertainment). Furthermore, since the performance of multicast communications is typically constrained by the worst channel conditions among users (e.g., those located farther away), PASS can enhance overall multicast performance by optimizing the activation positions of PAs to improve users' channel gains. To the best of the authors' knowledge, the design of PASS-enabled multicast communications remains unexplored. This provides the main motivation of this work.

Against the above background, this paper explores the benefits of PASS in multicast communications. In particular, we proposed a PASS-enabled multicast communication framework, where multiple PAs on a single waveguide send the broadcast information to multiple users located in the service region. We formulate the PA position optimization problem for the maximization of the multicast performance, i.e., maximizing the worst-case user performance, subject to the deployment constraints of PAs. To address this non-convex problem, a particle swarm optimization (PSO)-based algorithm is developed to determine the desired PA positions. Our numerical results verify the effectiveness of PASS with the significant performance gain over the conventional multiple-antenna transmission scheme.

\section{System Model and Problem Formulation}
In this section, we first introduce the considered PASS-enabled multicast communication system with its basic signal model. Then, we formulate the multicast performance maximization problem.

\begin{figure}[htbp]
  \centering
  \includegraphics[width=80mm]{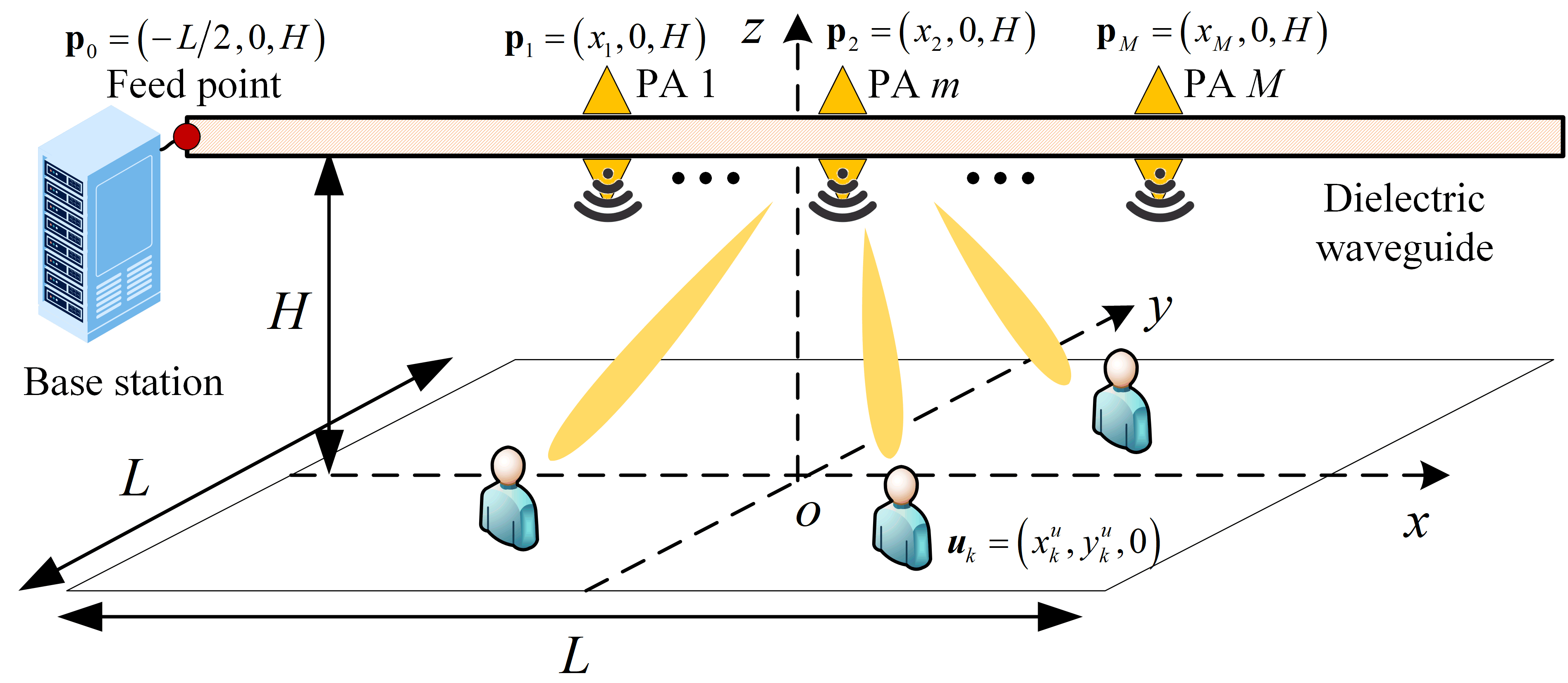}
  \caption{Illustration of PASS-enabled multicast communications.}
  \label{fig:MC}
\end{figure}
\subsection{System Model}
As shown in Figure~\ref{fig:MC}, we study a PASS-enabled multicast communication system, which consists of a base station (BS) connecting with one dielectric waveguide and $K$ single-antenna multicast users. There are $M$ PAs on the waveguide, which are activated using small dielectric particles (e.g., plastic pinches) to radiate signals from the waveguide into free space. Without loss of generality, a three-dimensional (3D) Cartesian coordinate system is considered. In particular, the users are located on the $x-y$ plane with a squared area of $L \times L$ meters ($m^2$), whose center is the origin point. Therefore, the location of the $k$th user is denoted by ${{\bf{u}}_k} = \left( {x_k^u,y_k^u,0} \right),\forall k \in \mathcal{K}  \buildrel \Delta \over = \left\{ {1,2, \ldots ,K} \right\}$. The waveguide with the length of $L$ is deployed at the height of $H$ and in parallel to the $x$-axis, staring from $\left( { - \frac{L}{2},0,H} \right)$ and ending at $\left( { \frac{L}{2},0,H} \right)$. As a result, the location of the $m$th PA is denoted by ${{\bf{p}}_m} = \left( {{x_m},0,H} \right),\forall m \in \mathcal{M}  \buildrel \Delta \over = \left\{ {1,2, \ldots ,M} \right\}$. 

Let ${{\bf{p}}_0} = \left( { - \frac{L}{2},0,H} \right)$ denote the feed point of the waveguide, where the signal transmits into the waveguide from the baseband of the BS. The in-waveguide channel vector between the feed point and the $M$ PAs is given by 
\begin{equation}
  \label{gx}
{\bf{g}}\left( {\bf{x}} \right) = {\left[ {{e^{ - j\frac{{2\pi }}{{{\lambda _g}}}\left( {{x_1} + {L \mathord{\left/
 {\vphantom {L 2}} \right.
 \kern-\nulldelimiterspace} 2}} \right)}},{e^{ - j\frac{{2\pi }}{{{\lambda _g}}}\left( {{x_2} + {L \mathord{\left/
 {\vphantom {L 2}} \right.
 \kern-\nulldelimiterspace} 2}} \right)}}, \ldots ,{e^{ - j\frac{{2\pi }}{{{\lambda _g}}}\left( {{x_M} + {L \mathord{\left/
 {\vphantom {L 2}} \right.
 \kern-\nulldelimiterspace} 2}} \right)}}} \right]^T},
\end{equation}
where ${\bf{x}} = \left[ {{x_1},{x_2}, \ldots ,{x_M}} \right]$ is the $x$-axis position vector of all $M$ PAs and ${\lambda _g}$ is the waveguide wavelength in a dielectric waveguide. Here, ${\lambda _g} = \frac{\lambda }{{{n_{eff}}}}$, where ${\lambda }$ the signal wavelength in the free space and ${n_{eff}}$ is the effective refractive index of a dielectric waveguide \cite{Ding}. Note that in \eqref{gx}, the propagation loss within the waveguide is omitted since its impact on the communication performance is shown to be negligible \cite{Kaidi}.

As the PAs are activated at positions that can have LoS links with users, the free-space channel vector between $M$ PAs and the $k$th user is given by
\begin{equation}
  \label{hkx}
{{\bf{h}}_k^H}\left( {\bf{x}} \right) = \left[ {\frac{{\sqrt \eta  }}{{{d_{k,1}}}}{e^{ - j\frac{{2\pi }}{\lambda }{d_{k,1}}}},\frac{{\sqrt \eta  }}{{{d_{k,2}}}}{e^{ - j\frac{{2\pi }}{\lambda }{d_{k,2}}}}, \ldots ,\frac{{\sqrt \eta  }}{{{d_{k,M}}}}{e^{ - j\frac{{2\pi }}{\lambda }{d_{k,M}}}}} \right],
\end{equation}
where ${\sqrt \eta  }  = \frac{c}{{4\pi {f_c}}}$. Here, $f_c$ is the carrier frequency and $c$ is the speed of light. ${d_{k,m}}$ represents the distance between the the $k$th user and the $m$th PA, which is given by
\begin{equation}
  \label{dkm}
{d_{k,m}} = \left| {{{\bf{u}}_k} - {{\bf{p}}_m}} \right| = \sqrt {{{\left( {x_k^u - {x_m}} \right)}^2} + v_k^2},
\end{equation}
where $v_k^2 = {\left( {y_k^u} \right)^2} + {H^2}$.

Let $s \in \mathbb{C}$ denote the normalized multicast signal at the baseband of the BS to be fed into the waveguide. Therefore, the received multicast signal at the $k$th user is given by
\begin{equation}
  \label{yk}
{y_k} = \sqrt {\frac{P}{M}} {{\bf{h}}_k^H}\left( {\bf{x}} \right){\bf{g}}\left( {\bf{x}} \right)s + {n_k},
\end{equation}
where ${n_k}\thicksim CN\left( {0,\sigma _k^2} \right)$ denotes the additive white Gaussian noise (AWGN) at the $k$th user. For simplicity, we assume that $\sigma _k^2 \buildrel \Delta \over = \sigma _0^2,\forall k \in {\mathcal K}$. $P$ is the total transmit power. In this work, the radiating signal model of PASS follows the equal power model proposed in \cite{PASS}, where the transmit power is equally allocated among $M$ PAs, i.e., ${\frac{P}{M}}$ per PA.

Accordingly, the received signal-to-noise-ratio (SNR) for the multicast signal at the $k$th user can be expressed as
\begin{equation}
  \label{SNR}
{\gamma _k} = \frac{P}{M}\frac{{{{\left| {{{\bf{h}}_k^H}\left( {\bf{x}} \right){\bf{g}}\left( {\bf{x}} \right)} \right|}^2}}}{{\sigma _0^2}} = \frac{{P\eta }}{{M\sigma _0^2}}\left| {\sum\limits_{m = 1}^M {\frac{{{e^{ - j\frac{{2\pi }}{\lambda }{d_{k,m}} - j\frac{{2\pi }}{{{\lambda _g}}}\left( {{x_m} + {L \mathord{\left/
 {\vphantom {L 2}} \right.
 \kern-\nulldelimiterspace} 2}} \right)}}}}{{{d_{k,m}}}}} } \right|^2.
\end{equation}

\subsection{Problem Formulation}
For multicast communications, the system performance is evaluated by the user achieving the lowest communication rate. Therefore, the performance of the considered PASS-enabled multicast communications is given by
\begin{equation}
  \label{MC_performance}
\overline \gamma   = \min \left\{ {{\gamma _k},\forall k \in {\mathcal K}} \right\}.
\end{equation}
Our goal is to maximize the multicast performance by optimizing the positions of $M$ PAs, ${\bf{x}}$. The corresponding optimization problem can be formulated as follows:
\begin{subequations}\label{problem}
\begin{align}
\mathop {\max }\limits_{\bf{x}} \;\;&\overline \gamma  \\
\label{range}{\rm{s.t.}}\;\;& - \frac{L}{2} \le {x_1} < {x_2} <  \ldots  < {x_M} \le \frac{L}{2},\\
\label{distance} & {x_{m + 1}} - {x_m} \ge \Delta ,\forall m \in {\mathcal M}/\left\{ M \right\}.
\end{align}
\end{subequations}
Here, \eqref{range} means that $M$ PAs are activated in a successive manner along the $x$-axis within $\left[ { - \frac{L}{2},\frac{L}{2}} \right]$. \eqref{distance} defines the minimum distance, $\Delta> 0$, of two adjacent PAs to avoid mutual coupling.

\section{PSO-based Solutions}

In this section, we will develop a PSO-based algorithm to address problem \eqref{problem}. To employ the PSO method, we first define $I$ initialization particles corresponding to the position of each PA as follows:
\begin{align}
	\mathbf{x}_i^{(0)}=[x^{(0)}_{i,1},x^{(0)}_{i,2},\cdots,x^{(0)}_{i,M}]^T,
\end{align}
where $x_{i,m}$ denotes the $x$-coordinate of the $m$th PA in the $i$th particle swarm, which satisfies $x_{i,m} \in [-\frac{L}{2},\frac{L}{2}]$.  The initial velocity of these particle swarms is defined as follows:
\begin{align}
	\mathbf{v}_i^{(0)}=[v^{(0)}_{i,1},v^{(0)}_{i,2},\cdots,v^{(0)}_{i,M}]^T.
\end{align}

Moreover, let $\mathbf{x}_{i,p}$ and $\mathbf{x}_{g}$ denote the personal best position of the $i$th particle and the global best position of all particles, respectively. Then, according to the iterative rules of the PSO algorithm \cite{PSO}, these parameters can be updated in each iteration as follows:
\begin{align}\label{velocity}
	\mathbf{v}^{(t+1)}_{i}\!\!=\omega\mathbf{v}^{(t)}_i\!+\!c_1\alpha_1(\mathbf{x}_{i,p}\!-\!\mathbf{x}^{(t)}_i)\!+\!c_2\alpha_2(\mathbf{x}_{g}\!-\!\mathbf{x}^{(t)}_i),
\end{align}
\begin{align}\label{position}
	\mathbf{x}^{(t+1)}_i=\mathbf{x}^{(t)}_i+\mathbf{v}^{(t+1)}_i,
\end{align}
where $t$ denotes the number of iterations. Here, $c_1$ and $c_2$ are the personal and global learning factors, respectively. $\alpha_1$ and $\alpha_2 \in [0,1]$ are random factors that improve search generalization. $\omega$ represents the inertia weight of the particle search. In particular, $\omega$ is a non-negative value, and we employ a dynamic $\omega$ to enhance the algorithm performance, defined as $\omega=\omega_{\max}-(\omega_{\max}-\omega_{\min})\frac{t}{T}$, where $\omega_{\max}$ and $\omega_{\min}$ are predetermined upper and lower bounds of $\omega$, respectively, and $T$ is the maximum number of iterations. Furthermore, to satisfy the constraint \eqref{range} of the optimization problem, the position of each PA must be adjusted after every iteration to stay within the defined boundaries as follows:
\begin{align}
	[\mathbf{x}^{(t)}_i]_m=\!\! \begin{cases}
	   -\frac{L}{2}, \quad  \textup{if} \quad	[\mathbf{x}^{(t)}_i]_m < -\frac{L}{2},\\ 
	   \frac{L}{2}, \quad  \ \ \ \textup{if} \quad	[\mathbf{x}^{(t)}_i]_m > \frac{L}{2},\\
	   [\mathbf{x}^{(t)}_i]_m, \quad  \ \  \textup{otherwise}.
	\end{cases}
\end{align}

The fitness of each particle is evaluated based on the unconstrained problem \eqref{problem}. In other words, given the PA positions in each iteration, the SNR of each user is calculated from \eqref{SNR}, and the smallest value is chosen as the reference for the fitness function. For convenience, we define it as $\Gamma(\mathbf{x}_i)$. Furthermore, in order to ensure constraint \eqref{distance}, we introduce a penalty factor $\xi>0$ to the fitness function and update it as follows:
\begin{align} \label{Fitness}
	\mathbb{F}(\mathbf{x}_i)=\Gamma(\mathbf{x}_i)-\xi\mathbb{P}(\mathbf{x}_i).
\end{align}
where $\mathbb{P}(\mathbf{x}_i)$ is a penalty function that counts the number of instances in which the current PA position violates the constraint \eqref{distance}, which is given by
\begin{align}
	\mathbb{P}(\mathbf{x}_i)=\frac{1}{2}\sum_{m=1}^{M}\sum_{n=1,n\ne m}^{M} \mathbb{I}(\|\mathbf{x}_{i,m}-\mathbf{x}_{i,n}\|<\Delta),
\end{align}
where $\mathbb{I}$ is an indicator function that takes the value of 1 when the condition in parentheses is true and 0 otherwise. Note that, to ensure the penalty term satisfies the distance limit between all pinching antennas, the penalty factor $\xi$ generally needs to be set to a large value. With the fitness evaluation conducted on each particle, their local and global best positions are improved until convergence. The exact processes are summarized in \textbf{Algorithm \ref{alg:A}}.

\begin{algorithm}[!t]\label{method2}
	\caption{Proposed PSO-based Algorithm to Solve Problem \eqref{problem}.}
	\label{alg:A}
	\begin{algorithmic}[1]
		\STATE {Set PSO related parameters $c_1$, $c_2$, $\alpha_1$, $\alpha_2$, $\omega_{\min}$, $\omega_{\max}$, $T$, and $\xi$.}
		\STATE Initialize the $I$ particles with position $\mathbf{x}^{(0)}_i$ and $\mathbf{v}^{(0)}_i$, the iteration index, $t=1$.
		\STATE Evaluate the fitness function $\mathbb{F}$ value for each particle using \eqref{SNR}.
		\STATE Obtain the local best position $\mathbf{x}_{i,p}=\mathbf{x}^{(0)}_{i}$ for each particle and the global best position $\mathbf{x}_{g} = \arg \max \{\mathbb{F}(\mathbf{x}^{(0)}_{1}),\cdots,\mathbb{F}(\mathbf{x}^{(0)}_{I}))\}$.
		\REPEAT 
		\STATE Update the inertia with $\omega=\omega_{\max}-\left(\omega_{\max}-\omega_{\min}\right)t/T$.
		\REPEAT 
		\STATE Update the velocity and position of particle $i$ according to \eqref{velocity} and \eqref{position}, respectively.\\
		\STATE Evaluate the fitness function $\mathbb{F}$ value for particle $i$ using \eqref{SNR} and update it according to \eqref{Fitness}.
		\STATE \textbf{if} $\mathbb{F}(\mathbf{x}^{(t)}_{i})> \mathbb{F}(\mathbf{x}_{i,p})$ \textbf{then}
		\STATE  \quad Update $\mathbf{x}_{i,p} \leftarrow \mathbf{x}^{(t)}_{i}$.
		\STATE  \textbf{end if}
		\STATE \textbf{if} $\mathbb{F}(\mathbf{x}^{(t)}_{i})> \mathbb{F}(\mathbf{x}_{g})$ \textbf{then}
		\STATE  \quad Update $\mathbf{x}_{g} \leftarrow \mathbf{x}^{(t)}_{i}$.
		\STATE  \textbf{end if}
		\STATE $i \leftarrow i+1$.
		\UNTIL $i>I$.
		\STATE $t \leftarrow t+1$.
		\UNTIL $t>T$.\\
		\STATE {Output} the optimal solutions $\mathbf{x}^*$.
	\end{algorithmic}
\end{algorithm}
\section{Numerical Results}
In this section, we provide numerical results to verify the effectiveness of PASS. The system parameters are set as follows: the number of users $K = 4$, $H = 3$ m, $f_c = 28$ GHz, $\Delta=\frac{\lambda}{2}$, $n_{eff}=1.4$, $P=1$ mW, and $\sigma^2=-90$ dBm. For the PSO algorithm, we set $c_1=c_2=1.5$, $\omega_{\max}=0.8$, $\omega_{\min}=0.2$, $T=1000$, and $\xi=1000$. For comparison, we consider the conventional multiple-antenna transmission as a baseline, where the $M$-antenna BS is fixed at $\left( { -\frac{L}{2},0,H} \right)$. The achieved multicast rate is calculated by ${\log _2}\left( {1 + \overline \gamma  } \right)$.

\begin{figure}[t]
	\centering
	\includegraphics[width=80mm]{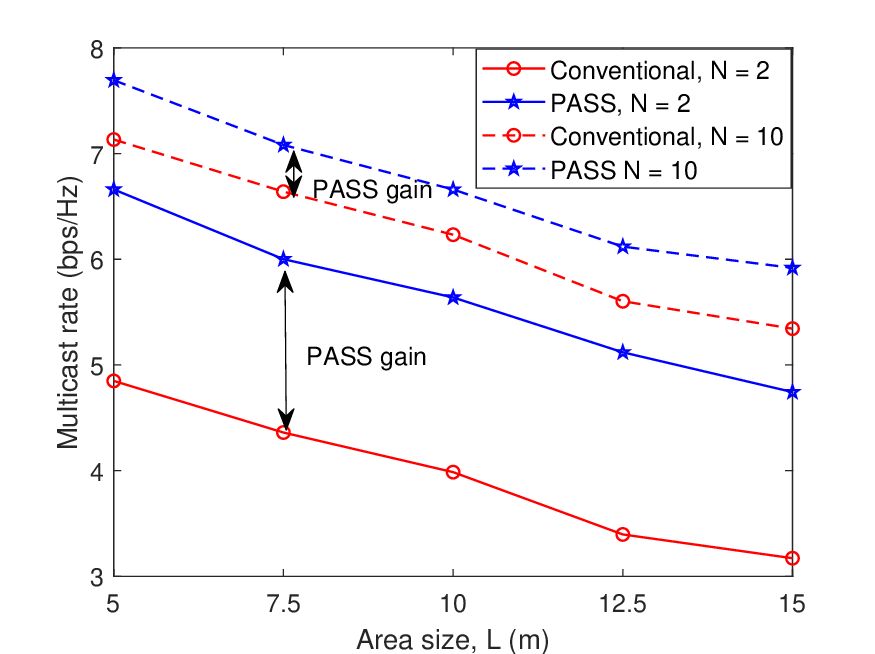}
	\caption{Multicast rate versus the area size, $L$.}
	\label{fig:area}
\end{figure}

In Figure \ref{fig:area}, we study the achieved multicast rate versus the area size, $L$, where $M=2$ and $M=10$. It can be observed that PASS significantly outperforms the conventional multiple-antenna BS, especially for a limited number of $M=2$. This is expected since by dynamically adjusting the positions of PAs, PASS can avoid long-distance path loss and serve the users via the short-range LoS link. Moreover, the performance gain of PASS decreases with the increase of $M$. This is because a larger $M$ provides higher spatial degrees-of-freedom for the conventional multiple-antenna BS. This, however, leads to high hardware costs since each antenna needs a dedicated costly and power-hungry radio-frequency chain. It can be also observed that the multicast rate achieved by PASS and conventional multiple-antenna BS both decrease with the increase of $L$. This is because a larger area and randomly distributed users in general increase the communication distance for the two schemes, i.e., higher path loss.

\begin{figure}[t]
	\centering
	\includegraphics[width=80mm]{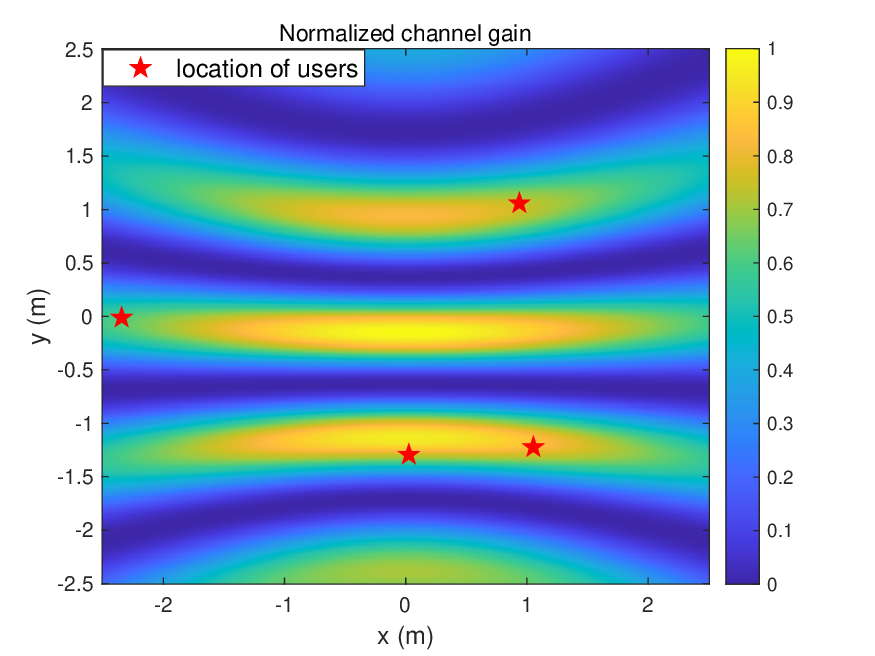}
	\caption{Normalized channel gain achieved by PASS over the area, where $M=2$ and $L=5$.}
	\label{fig:channel}
\end{figure}

In Figure \ref{fig:channel}, we further study the normalized channel gain achieved by PASS over the area, where $M=2$ and $L=5$. Here, $\bigstar$ denotes the locations of 4 users.  Thanks to the near-field channel models in \eqref{gx} and \eqref{hkx}, it can be observed that PASS can achieve the beamfocusing pattern with only 2 PAs and try to focus the signal's energy on the locations of different users. This again confirms the effectiveness of PASS.

\section{Conclusions}
A novel PASS-enabled multicast communication framework has been proposed, which employs multiple PAs on a single waveguide to deliver the broadcast information to served users. A multicast performance maximization problem was formulated to optimize the position of each PA, which was solved by the developed PSO-based algorithm. Numerical results confirmed the effectiveness of PASS and showed that the proposed PASS scheme outperforms the conventional multiple-antenna transmission scheme.

\bibliographystyle{IEEEtran}
\bibliography{refs}

\end{document}